# Status of Nano-ARPES endstation at BL07U of Shanghai Synchrotron Radiation Facility

Han Gao[1], Hanbo Xiao[1], Feng Wang[1], Fangyuan Zhu[2], Meixiao Wang[1], Zhongkai Liu[1], Yulin Chen[1], Cheng Chen[1]

[1]*School of Physical Science and Technology, ShanghaiTech Laboratory for Topological Physics, ShanghaiTech University, Shanghai 201210, China.*
[2]*Shanghai Synchrotron Radiation Facility, Shanghai Advanced Research Institute, Chinese Academy of Sciences, Shanghai 201204, China.*

## Introduction

Angle-resolved photoemission spectroscopy (ARPES) is a critical technique in modern condensed matter physics research, known for its ability to directly visualize the electronic structure of materials. By resolving the energy and momentum information of the photoelectrons excited from materials using ultraviolet/X-ray radiation, ARPES provides direct insights into the band structure and interactions within a quantum system. This technique has made essential contribution in the exploration of many complex material systems, such as high $T_c$ superconductors, topological insulators, strong correlated electron systems[1]. However, limited by the spot size of incoming beam, typically on the order of tens to hundreds of microns, conventional APRES experiments mostly focused on the cleaved single crystals or the epitaxial thin film, which offer large and uniform areas for detection. Fortunately, recent advancements in X-ray focusing techniques have led to the development of ARPES systems with enhanced spatial resolution, known as Nano-ARPES technique. This innovation makes it possible to investigate various systems with limited geometrical sizes, such as quasi-1D systems, 2D heterostructures/Moiré systems, *in-operando* tunable devices, material with spatial phase separation etc.

Nano-ARPES endstation, turning into user-operation mode in 2021, was constructed on the main branch at BL07U beamline ($S^2$-line) of Shanghai Synchrotron Radiation Facility (SSRF), which also includes a second branch consisting of a spin-resolved ARPES endstation and a vector magnetic field/high magnetic field X-ray Magnetic Circular Dichroism (XMCD) endstation. NanoARPES endstation adopts the Zone plate optical systems, which focus the incident synchrotron radiation down to ~200nm, with full polarization control and optimized around 98eV. The piezo-motor-driven sample stage provides high spatial precision with 5 degrees of motion, and is equipped with 4 contacting electrodes, providing the capabilities of *in-operando* operations. The sample temperature covers the range from 14K to room temperature, and the system is equipped with a DA30 hemispherical electron analyzer with a total energy resolution around 10meV. To facilitate the research on 2D materials, the endstation is equipped with a state-of-the-art ultra-high vacuum (UHV) 2D fabrication system. This system allows for the *in-situ* preparation of artificially stacked 2D heterostructure without surface contamination, a critical factor for ARPES measurements. Detailed information about this Nano-ARPES endstation will be introduced in the following.

**Technical specifications**

Figure 1a presents a schematic of the $S^2$ beamline, where the beam is switched between two branches by the CM3/PM4 plane mirror (see photo in Fig. 1b). The beamline is equipped with three Plane Grating Monochromators (PGMs), including G400, G800 and G1200 lines, while the Nano-ARPES system primarily utilizes the former two, due to their relative higher incoming photon flux (~$3\times10^{12}$ phs/s). The photon energy of the beamline ranges from 50eV to 2000eV, while the NanoARPES typically operates between 85eV-105eV, constraint by the use of Zone plate. The angle of incident beam relative to the electron analyzer is 60° in normal operating conditions.

Nano-ARPES endstation (shown in the photo in Fig. 2c) mainly consists of four UHV chambers: these include a fast-in loadlock chamber for rapid sample insertion, an interface chamber for sample transfer and storage, a preparation chamber for sample treatment (such as annealing and sputtering), and a main chamber for NanoARPES measurement. Unlike traditional ARPES systems, where the sample stage/manipulator is rotatable to collect photoelectrons at different emission angles while the electron analyzer and main chamber remain stationary, the NanoARPES endstation is designed in the opposite manner. To achieve a high-level of spatial stability, the optical/sample stage is seated on a heavy granite base and remains stationary during experiments, while the electron analyzer rotates instead. Therefore, the entire system, including all four UHV chambers and analyzer, is built on a large ball bearing with a diameter of 1.9m, allowing for a rotation range of ±15°. A differential pumping rotation platform (DPRF) bridges the rotating parts with the sample stages and granite base to maintain a high level of vacuum, and a rectangular-flexible-welded bellow is used on the beam-entrance connection, accommodating for the chamber rotation.

Figure 1d illustrates the details of the optical and sample stages, mainly consisting of 11-axis piezo-motors with high spatial precisions. These include 3 translational motions for Fresnel Zone Plates (FZP) and Order Sorting Aperture (OSA) respectively, and 5 degree of motions (3 translation plus polar and tilt rotations) for the sample stage. Four contacting electrodes are integrated into the sample stage, enabling *in-operando* operation during Nano-ARPES measurement. These operations include electrostatic gating (as demonstrated later in Fig. 3), current application, and other functions depending on the specific sample design. The base temperature of the measurement is 14K, and the pressure is maintained at better than $1.5 \times 10^{-10}$ mbar.

The focusing of the incident beam is achieved through the combined use of FZP and OSA, as illustrated in Fig. 2a. The effective diameter of the FZP equipped is around 2.5mm (12550 zones) with an operational center energy of 100eV and a working distance of 10mm. Incident photobeam, positioned 4.75m downstream from the beamline exit slit, forms a 3.51 mm (W) × 3.77 mm (H) beam spot on the FZP, within 4σ of gaussian distribution. The first-order diffraction of the beam is selected through the combined effects of center stop and OSA, typically using a pinhole size of 0.5 mm, as shown in Fig. 2b. A multichannel plate (MCP) is installed downstream, behind the sample stage, and is typically used before experiments to examine the collimation of optical components with the incident photon beam. To calibrate the spatial resolving power of the system, we conducted a real-space scanning photoemission experiment on a standard test sample, which consists of pre-patterned gold wires with different spacing on a silicon substrate. The Scanning Electron Microscopy (SEM) image and real-space photoemission map are presented in Fig. 2(c, e), along with integrated intensity curves in Fig. 2(d, f). Here the gold wires are clearly resolved along both the vertical and horizontal directions. Considering the 60° angle between the incident beam and the sample's normal direction, as well as the thickness of the gold wires (50nm), the real spot size is estimated to be around 200 x 500nm. This result can be further improved with fine tuning of the incident beam before FZP and the use of a calibration sample with sharper edges.

Lastly, the endstation uses a standard DA30 electron analyzer from ScientaOmicron, equipped with an electron deflector that collects electrons with an additional ±15° on top of the chamber rotation range. The overall energy resolution is calibrated to be around 10meV at 91eV photon energy.

**NanoARPES measurements on 2D devices**

The unique spatial resolving capabilities of the NanoARPES technique enable investigations on the electronic structure of various material systems with limited geometrical sizes. Examples include cleaved single crystals with rough surfaces and small domains, spatial phase separation systems, artificially fabricated 2D devices via mechanical exfoliation, *in-operando* tunable devices, quasi-1D/0D systems etc. Particularly, recent advances in 2D materials provide versatile platforms for investigating many intriguing physics phenomena[2,3]. These 2D layers on one hand exhibit exotic emergent phenomena themselves, such as superconductivity, magnetism, and quantum spin hall effect etc. On the other hand, by stacking identical 2D layers with a rotational misalignment, or dissimilar 2D layers with lattice mismatch, a 'Moiré system' can be created. In these systems, an enlarged Moiré superlattice (usually on the order of 10nm) leads to the flatness of the electronic bands, resulting in the emergence of strong electron correlation effects and topological properties[4,5,6]. Many intriguing electronic states have been realized thereafter, even including those that have not yet been discovered in the bulk crystals, such as the recently discovered fractional quantum anomalous hall states[7,8].

The exploration of the electronic structure of these quantum systems can provide essential insights for unraveling the intricate interplay between lattice, moiré potential and electron correlations. The small geometrical size (typically on the order of microns), limited by the mechanical exfoliation technique, urges the need of NanoARPES technique. In the following, we will first present some basic measurement of the electronic structure of exfoliated 2D thin films, as well as the *in-operando* tuning of the electronic states through electrostatic gating. Moreover, an advanced UHV sample fabrication method, compatible with the NanoARPES measurement to overcome the air instability of these 2D layers, will also be demonstrated.

Figure 3a illustrates a typical sample setup, where the gate-tunable bilayer graphene device is fabricated on the silicon substrate with prepatterned gold contacts. The sample geometry is shown in Fig. 3b, where the graphite back gate is isolated from the sample with a hexagonal boron nitride (hBN) flakes, as a dielectric layer. The graphene sample is grounded through one of the electrodes to ensure electrical conductivity during the photoemission experiment, and the graphite back gate is connected with a Keithley voltage source through another electrode. It is worth noting that the four electrodes on the sample stage are isolated from each other, and can be independently connected to different voltage/current sources or the ground. Integrating the photoemission intensities near the Fermi level, the real-space NanoARPES scanning (Fig. 3c(ii)) allows different sample regions to be distinguishable based on their different conductivity (Fig. 3c(i)). The characteristic band feature of bilayer graphene, the Dirac cones at the K point of Brillouin Zone (BZ), can be visualized in the scanning of the energy-momentum space at one particular sample position (Fig. 3 c(iii)). Applying a positive gate voltage can further tune the chemical potential of the system, bringing down the previously unoccupied conduction band into visualization[9]. Similarly, the exfoliated and artificially stacked twisted bilayer $WSe_2$ device, can also be precisely located in a NanoARPES measurement (Fig. 3d), and the characteristic band feature of $WSe_2$ around the $\Gamma$ point of BZ can be obtained, showing the split bands originated from inter-layer hybridization. Recently, electrostatic gating on these semiconducting transition metal dichalcogenides (TMDs) devices, has also been tested successful[10,11], paving the way for further investigation on their exotic electronic structure. It is worth noting that more delicate device designing is required to overcome the difficulties of electrical contact between electrodes and these semiconducting TMD flakes.

**UHV 2D fabrication system**

Limited by the mean free path of the excited photoelectrons, ARPES is a surface technique and is extremely sensitive to the surface condition of the samples. Traditional APRES measurements cleave the bulk single crystals inside the UHV chamber (typically less than $1\times10^{-10}$ mbar) to expose ultra clean surfaces without any contaminations. However, the 2D devices constructed through mechanical exfoliation and transfer, are usually prepared ex-situ, (e.g. in air and/or in the glove box) with multiple steps. Although we can apply complementary surface cleaning techniques, including ex-situ AFM cleaning, *in-situ* high-temperature annealing, etc, the surface quality of the samples still cannot be recovered to the level of *in-situ* cleaved single crystals.

To obtain 2D devices with excellent surface quality, we constructed a state-of-the-art UHV 2D fabrication system at the beamline (the design closely follows the work reported in Ref. 12), compatible with the NanoARPES measurement. The schematic diagram of the sample preparation using this UHV system is illustrated in Fig. 4a. The 2D samples are firstly thinned down to thin layers in air, through mechanical exfoliation using Kapton tape (compatible with UHV). Then, these tapes with samples are face-down attached to a silicon substrate, which is pre-deposited with 2nm Cr and 3nm Au through standard electron-beam evaporation. The second exfoliation of the tape from this substrate is processed within UHV, where clean sample surfaces can be obtained. The stickiness of the Au layer ensures the success of the exfoliation, and its conductivity guarantees the charge supply during ARPES measurement. To identify the spatial locations of the exfoliated thin films, the system is equipped with a long-working-distance microscope to observe the geometric morphology of the sample, where the thickness of the flakes can be judged by the color contrast. Moreover, to realize sample transfer and electrode fabrication all within UHV chamber, we construct a moving mechanism with 6 piezo-motors, including 3 translations for the transferring stage and 2 translations and 1 rotation for the sample stage. The temperature is controlled through a ceramic heating plate and a thermocouple installed under the sample stage, and the base pressure of the chamber is kept under $1\times10^{-9}$ mbar. The fabricated 2D devices are currently transferred to the NanoARPES system via vacuum suitcase, and we plan to directly connect the UHV 2D fabrication system to the NanoARPES system in the near future, further reducing the effort of sample transfer.

Some preliminary results obtained using these device fabrication systems are illustrated in Fig. 4(c-d). Here, a 2H-WS$_2$ flakes was exfoliated onto the Au substrate for testing purposes. Real-space photoemission scanning (Fig. 4c(ii)) successfully identified the thin flakes, as have been previously located in the corresponding microscopic image (Fig. 4c(ii)). Figure 4d shows characteristic dispersion cuts around the Γ point in BZ for both bulk and bilayer 2H-WS$_2$. These cuts reveal the characteristic band dispersions, including a split valence band top for the 2L sample and a broadened valence band top for the bulk sample. Besides these air stable samples like 2H-WS$_2$ that can be fabricated with traditional 2D transfer technique in air, this UHV system will be more useful for samples, whose surfaces degrade quickly in air — a common scenario for many materials.

**Summary**

After approximately six years of construction and commissioning, BL07U at the SSRF has become the first Nano-ARPES endstation in China, beginning its user-operation mode in 2021. This endstation offers exceptional spatial resolution (less than 200 nm) as well as decent energy and momentum resolutions. The inclusion of 4 contacting electrodes allows for *in-operando* tuning during ARPES measurements, enhancing the dynamic analysis capabilities. Additionally, the ultra-high vacuum (UHV) 2D fabrication system enables *in-situ* device fabrication, ensuring ultra-clean sample surfaces without contamination. As NanoARPES technique has already started to contribute essential insights in the recent studies of 2D materials, for instance the visualization of the flat band in magic angle twisted bilayer graphene (MATBG) systems[13-16], we sincerely hope that this new endstation in China will significantly broadens the exploration of electronic structures in more geometrically constrained material systems.

**Acknowledgments**

This work was supported by the Shanghai Municipal Science and Technology Major Project (grant no. 2018SHZDZX02). Y.L.C. acknowledges the support from the Oxford-ShanghaiTech collaboration project. Z. K. Liu acknowledges the support from the National Natural Science Foundation of China (92365204, 12274298) and the National Key R&D program of China (Grant No. 2022YFA1604400/03). F.Y. Zhu acknowledges the support from the National Natural Science Foundation of China (No. 52032005).

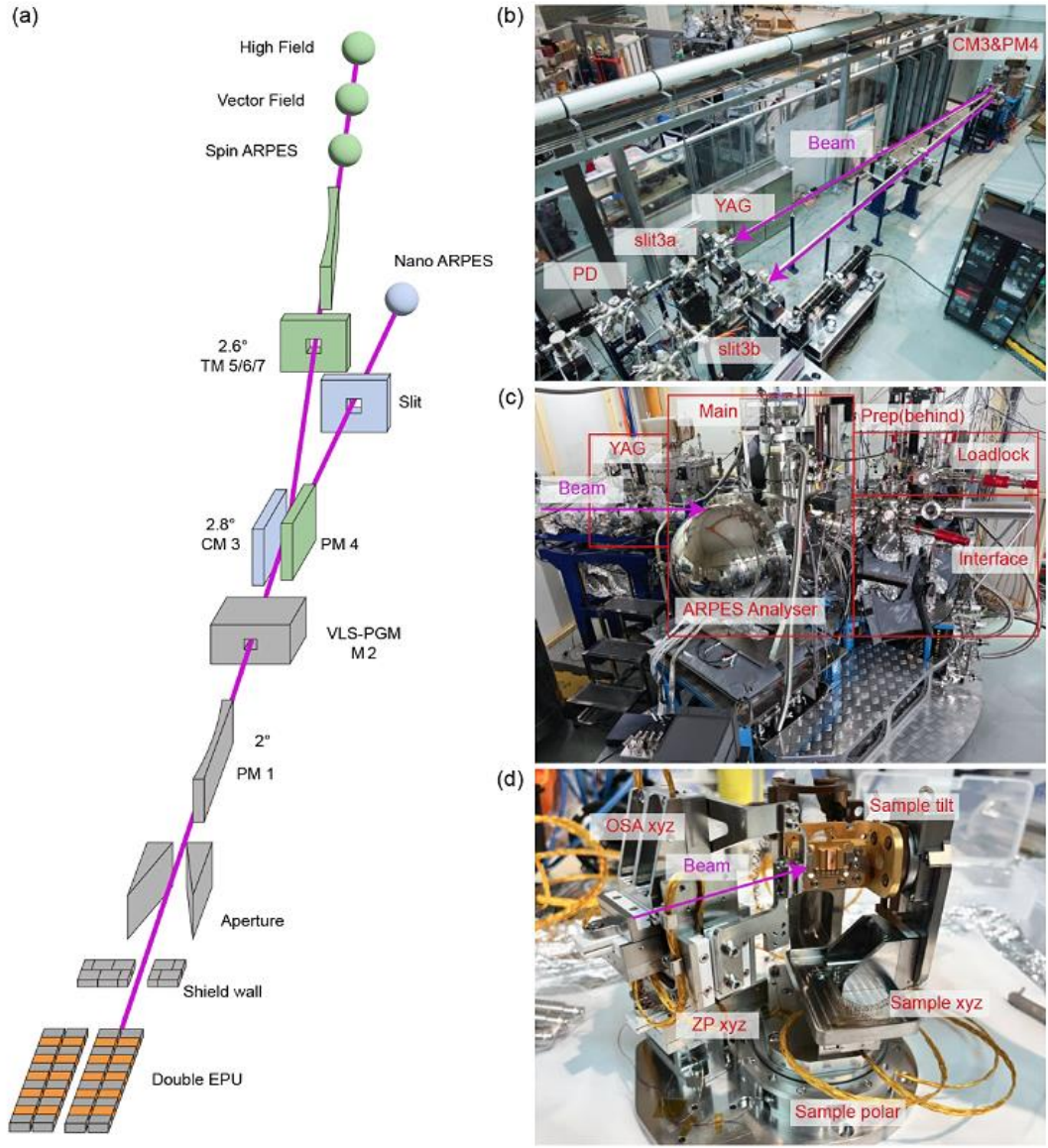

**Fig.1 Overview of BL07U at SSRF, and the Nano-ARPES endstation.** (a) Schematic of the optical setup at BL07U. (b) Photo that shows the layout of beamline from CM3&PM4 to NanoARPES endstation in (a). Purple arrow marks the beam direction. PD: photo diode. (c) Overview photo of NanoARPES system, including four UHV chambers and hemispherical electron Analyzer. YAG crystal at the front is used to diagnose the beam position. (d) Photo of 11-axis piezo-motor stage in the Main chamber of NanoARPES system. ZP: zone plate. OSA: order sorting aperture.

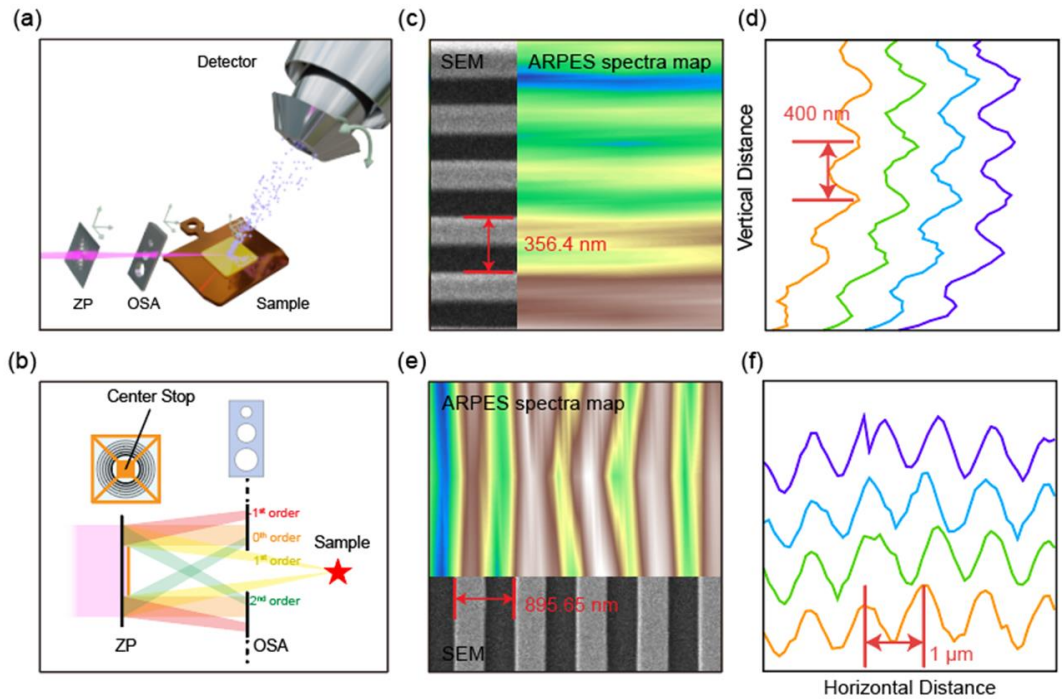

**Fig.2 Focusing principle and spatial resolution of NanoARPES.** (a) Illustration of NanoARPES measurement. (b) Schematic of zone plate optical system. ZP: zone plates. OSA: order sorting aperture. 1st order diffraction beam is selected through combined use of center stop and OSA. (c) SEM image and ARPES real-space photoemission intensity map of the test samples with pre-patterned gold wires. (d) Photoemission intensity curves of (c), illustrating the spatial resolution of NanoARPES along vertical direction. (e-f) same as (c-d), but for horizontal direction.

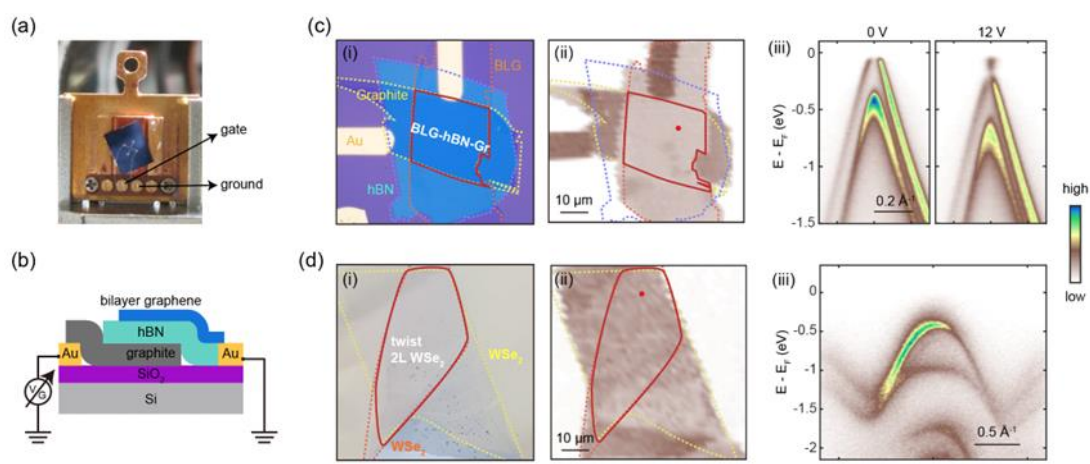

**Fig.3 NanoARPES measurements on 2D devices.** (a) Photo of four electrodes sample holder. The four pins on the holder are isolated from each other and can be connected to external voltage sources. Sample device is connected with the pins through wire bonding. (b) Schematic of an electrostatic gating device. (c) (i) Optical image, (ii) real-space photoemission intensity map of the sample. The red solid line marks the sample region and the red dot marks the measurement position. (iii) ARPES spectra with 0V and 12V gating voltage at the K point of bilayer graphene. (Reproduced from Ref. 9) (d) (i-iii) same as (c)(i-iii) but for twist bilayer $WSe_2$ on highly oriented pyrolytic graphite (HOPG).

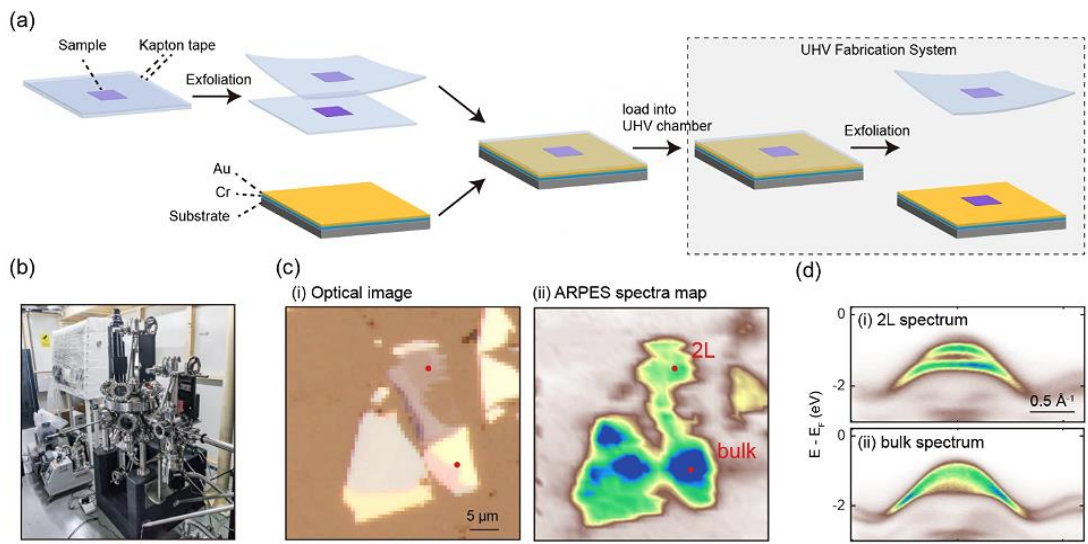

**Fig.4 *In-situ* sample preparation technique with UHV fabrication system.** (a) Schematic diagram of sample fabrication method inside UHV chamber. (b) Photo of UHV fabrication system at BL07U, SSRF. (c)(i) Optical image, (ii) real-space photoemission intensity map of 2H-WS$_2$ on Au/Si (SiO$_2$). The red dots mark the measurement position of (d). (d) ARPES spectra of (i) 2L and (ii) bulk 2H-WS$_2$.